# An Inorganic Liquid Crystalline Dispersion with 2D Ferroelectric Moieties

Ziyang Huang[1][†], Zehao Zhang[1][†], Rongjie Zhang[1], Baofu Ding[1,2][*], Liu Yang[3], Keyou Wu[1], Youan Xu[1,4], Gaokuo Zhong[5], Chuanlai Ren[5], Jiarong Liu[1], Yugan Hao[1], Menghao Wu[3], Teng Ma[6], and Bilu Liu[1][*]

[1] Shenzhen Graphene Centre, Tsinghua−Berkeley Shenzhen Institute and Institute of Materials Research, Tsinghua Shenzhen International Graduate School, Tsinghua University, Shenzhen 518055, China.

[2] Institute of Technology for Carbon Neutrality, Faculty of Materials Science and Engineering, Shenzhen Institute of Advanced Technology, Chinese Academy of Sciences, Shenzhen 518055, China.

[3] School of Physics and Institute for Quantum Science and Engineering, School of Chemistry and Institute of Theoretical Chemistry, Huazhong University of Science and Technology, Wuhan 430074, China.

[4] Xi'an Research Institute of High Technology, Xi'an 710025, China.

[5] Shenzhen Institute of Advanced Technology, Chinese Academy of Sciences, Shenzhen 518055, China.

[6] Department of Applied Physics, Hong Kong Polytechnic University, Hung Hom, Kowloon, Hong Kong 999077, China.

[†] These authors contributed equally to this work.

[*] Corresponding authors: bf.ding@siat.ac.cn (B.D.); bilu.liu@sz.tsinghua.edu.cn (B.L.)

**Abstract**

Electro-optical effect based liquid crystal devices have been extensively used in optical modulation techniques, in which the Kerr coefficient reflects the sensitivity of the liquid crystals and determines the strength of the device operational electric field. The Peterlin-Stuart theory and the O'Konski model jointly indicate that a giant Kerr coefficient could be obtained in a material with both a large geometrical anisotropy and an intrinsic polarization, but such a material is not yet reported. Here we reveal a ferroelectric effect in a monolayer two-dimensional mineral vermiculite. A large geometrical anisotropy factor and a large inherent electric dipole together raise the record value of Kerr coefficient by an order of magnitude, till $3.0\times10^{-4}$ m V$^{-2}$. This finding enables an ultra-low operational electric field of $10^2 \sim 10^4$ V m$^{-1}$ and the fabrication of electro-optical devices with an inch-level electrode separation, which is not practical previously. Because of its high ultraviolet stability (decay <1% under ultraviolet exposure of 1000 hours), large-scale, and energy-efficiency, prototypical displayable billboards have been fabricated for outdoor interactive scenes. The work provides new insights for both liquid crystal optics and two-dimensional ferroelectrics.

**Main**

Liquid crystal (LC) devices based on electro-optical effects have achieved a huge success in the information era, where the electric field alters the alignment of anisotropic LC molecules, giving rise to a change of their observed optical properties and achieving light modulation[1-4]. Electro-optical Kerr effect, firstly reported in 1875 by John Kerr, is one of a typical representative[5,6]. The Kerr coefficient that describes a quadratic relationship between induced birefringence and electric field strength reflects not only the strength of electro-birefringence effect, but also the sensitivity of the alignment of LC molecules in response to the electrical stimulus[7,8]. However, the values of Kerr coefficient are usually within the range of $10^{-11} \sim 10^{-7}$ m V$^{-2}$ for commercial LCs, and all require an operational electric field of $>10^6$ V m$^{-1}$ for LC electro-optical devices[7,8]. Such a high field results in a heavy reliance on sophisticated micromachining technology to pattern costly transparent electrodes on the optical path, where the electrodes need to be separated in micrometer-scale space with high uniformity[3]. In the meantime, it may lead to electrophoresis of ion species under high fields, and the consequent operational instability hinders the applications of lyotropic or charged LC systems in various disciplines with an electric driven demand[9,10]. These limitations make the classical LC devices technically and economically unacceptable for large-scale or outdoor uses. Therefore, low Kerr coefficients have become the bottleneck for extending the applications of LC electro-optical devices.

To obtain a large Kerr coefficient, the Peterlin-Stuart theory and the O'Konski model have been studied since 1940s, and from these scientists have proposed that materials with an intrinsic polarization and a large geometrical anisotropy are promising[11-13]. Organic ferroelectric LCs and inorganic LCs formed from low-dimensional materials have been developed to meet these criteria[14-20]. Organic ferroelectric LCs with a macroscopic polarization order have more sensitive response than their traditional counterparts[21,22], but their molecular scale with a small geometrical anisotropy limits further improvement of the Kerr coefficient. Two-dimensional (2D) materials with a micrometer-scale length and an atomic level thickness are believed to have the largest geometrical anisotropy among all the other materials in dispersion[23]. As the preparation techniques of 2D materials become fully developed, the increase of the Kerr coefficient is especially challenging, because the geometrical anisotropy approaches the upper limit[24]. In this regard, ideal Kerr media, namely inorganic LCs or LC-like systems based on 2D materials with both an intrinsic polarization and a large geometrical anisotropy, are urgently needed, while such LCs have not been reported.

Here we show such a 2D vermiculite (VMT) liquid crystalline dispersion with an anomalously large Kerr coefficient of $3.0\times10^{-4}$ m V$^{-2}$, which is an order of magnitude higher than all known media. The exfoliated 2D VMT has both a large geometrical anisotropy factor over 1500 and an intrinsic ferroelectricity, and these properties jointly contribute to the giant Kerr coefficient. For 2D VMT liquid crystalline dispersion, we establish a relationship among Kerr coefficient $K$, intrinsic polarization $P$, and geometrical anisotropy factor γ, namely, $K \propto P^2\gamma^4$, which is experimentally verified. Thanks to the giant Kerr coefficient, it is now possible for us to demonstrate a large-area prototypical display with low-energy consumption for potential outdoor use.

An aqueous 2D VMT dispersion was prepared by exfoliating bulk layered minerals by a cation-exchange method (see Methods and Supplementary Fig. 1). During this process, the interlayer ions of VMT were replaced by cations in solution (e.g., $Na^+$, $Li^+$). The interlayer spacing of VMT expanded, so that the interlayer van der Waals force was weakened[20,25,26]. Hence, we were able to produce VMT monolayers and their dispersion (Supplementary Fig. 2). Note that, the deionization treatment to decrease the ionic strength of the dispersion is a prerequisite in this study, because the ion-induced electrostatic screening and/or electrophoresis can lead to either the weakened impact of electric field on the 2D VMT or large leakage current (Supplementary Fig. 3). The 2D VMT dispersion had a greenish brown color and exhibited colloidal behavior, as evidenced by its Tyndall effect (Supplementary Fig. 4a-b). A Zeta potential of -48.1 mV indicated that the 2D VMT had a negatively charged surface (Supplementary Fig. 4c), where the repulsive electrostatic force between adjacent platelets prevents their aggregation and consequently imparts stability to the dispersion. When placed under crossed polarizers, we observed that 2D VMT dispersion underwent a phase transition from an isotropic phase to biphasic and anisotropic ones with an increasing volume fraction at a static state (Fig. 1a-c). When it is shaken by hand, diluted 2D VMT dispersion exhibited flow-induced birefringent textures with a brush shape (Supplementary Fig. 5a-c), where the texture was similar with other lyotropic liquid crystalline dispersion[16-18,27]. The dark brushes indicated that the dispersed 2D VMT platelets were temporarily parallel to one of the crossed polarizer axes, making corresponding area to be extinction. In contrast, the 2D VMT platelets collectively oriented in an intermediate direction in the bright regions. More fringes with birefringence colors (e.g., red, purple, and green ones) appeared as the volume fraction increased. These temporary ordered domains and birefringent textures disappeared both globally and locally within several minutes, and these dilute 2D VMT dispersion returned back to an optically isotropic state after shaking (see Supplementary Fig. 6 and Supplementary Fig. 7). Without a flow, a 2D VMT dispersion with a volume fraction of 0.40 vol% (or higher) showed a spontaneous Schlieren texture, corresponding to the anisotropic nematic phase (see both Fig. 1c and Supplementary Fig. 5d). These results indicated that 2D VMT dispersion had liquid crystalline properties. Small angle X-ray scattering and rheological tests were performed to observe the phase transition behavior as well. The elliptical small angle X-ray scattering pattern and the resultant profile showed that the 2D VMT dispersion with a volume fraction of 1.2 vol% became anisotropic, where the long axis of the elliptical pattern was coplanar with the normals of 2D VMT platelets (Supplementary Fig. 8). In the rheological test, 2D VMT dispersion showed volume-fraction-induced thickening effect and shear-induced thinning effect, which agreed with the lyotropic liquid crystalline nature of 2D VMT dispersion (Supplementary Fig. 9).

Figure 1d-g presents the electro-optical birefringence response of 2D VMT liquid crystalline dispersion. A cuvette filled with 2D VMT liquid crystalline dispersion appeared black in the absence of an electric field, because the disordered VMT platelets were optically isotropic, giving rise to optical extinction under crossed polarizers (Fig. 1d). It turned bright after applying a transverse electric field of $1.0 \times 10^3$ V m$^{-1}$ with a frequency of 10 kHz, due to the transmission of partial light from the back light source (Fig. 1e), while an electric field of about $10^6$ V m$^{-1}$ is usually needed for organic LCs. It indicates a drop of operational electric field strength by three orders of magnitude. This was also observed by polarized optical microscope (Fig. 1f-g). According to the Malus' Law, such an electro-optical switch depends on the birefringence $\Delta n$ controlled by LC alignment[28,29], and the electric field induced alignment can be confirmed by polarization-dependent transmittance[30]. In general, the 2D VMT liquid crystalline dispersion was initially optically isotropic due to the random orientation of 2D VMT platelets. When the platelets are aligned by external stimulus, an anisotropic

absorption occurs (Supplementary Fig. 10). The transmitted light intensity with the incident light polarized perpendicular to the electric field was higher than that parallel to the electric field (Fig. 1h). Besides, some textures that parallel to the transverse electric field was observed as well (Supplementary Fig. 11). These results collectively indicate that the 2D VMT platelets aligned in parallel to the electric field $E$, i.e., their normals are perpendicular to $E$.

As $E$ approaches 0, $\Delta n$ and $E$ theoretically satisfy the relationship of $\Delta n = K\lambda E^2$ ($K$ is the Kerr coefficient and $\lambda$ the incident wavelength), consistent with the experimental observation (Fig. 1i). Here, $\Delta n$ was determined by calculating the polarization parameters of azimuth angle and ellipticity (see Methods, Supplementary Text 1, Supplementary Table 1, Supplementary Fig. 12, and Supplementary Fig. 13). For 2D VMT liquid crystalline dispersion with a volume fraction of 0.08 vol%, the Kerr coefficient was determined to be $3.0\times10^{-4}$ m V$^{-2}$ (Fig. 1j). It is worth noting that the Kerr coefficient of 2D VMT liquid crystalline dispersion is one order of magnitude higher than that of all known Kerr media, including organic blue phase LCs, organic ferroelectric LCs, and all inorganic nanomaterial LCs (see Fig. 1l and Supplementary Table 2). Theoretically, the Peterlin-Stuart theory gives the basic relationship between the birefringence, optical anisotropy factor $\Delta g$ of LC materials and orientational order parameter[11,12]. Besides, the O'Konski model illustrates that $\Delta\alpha$, the anisotropy of excess electrical polarizabilities of LC unit beyond the electrical polarizabilities of dispersant, and $\mu$, the inherent electric dipole of each LC unit, jointly determine the orientational order parameter[13]. In simple, we showed that the Kerr coefficient was positively correlated with $\Delta g$, $\Delta\alpha$ and $\mu$ (Supplementary Text 2). Measurements of saturated birefringence and the frequency dependence of the Kerr coefficient experimentally gave that $\Delta g$, $\Delta\alpha$ and $|\mu|$ for 2D VMT liquid crystalline dispersion were $-2.3\times10^{-12}$ C$^2$ J$^{-1}$ m$^{-1}$, $-9.3\times10^{-27}$ F m$^2$ and $1.7\times10^{-23}$ C m ($5.1\times10^6$ Debye), respectively (see Fig. 1k, Supplementary Text 3, Supplementary Text 4, and Supplementary Fig. 14). The negative values of $\Delta g$ and $\Delta\alpha$ demonstrated an in-plane easy axis. Surprisingly, there existed an inherent electric dipole $|\mu|$ of $1.7\times10^{-23}$ C m, which possibly indicated the ferroelectric nature of 2D VMT platelets.

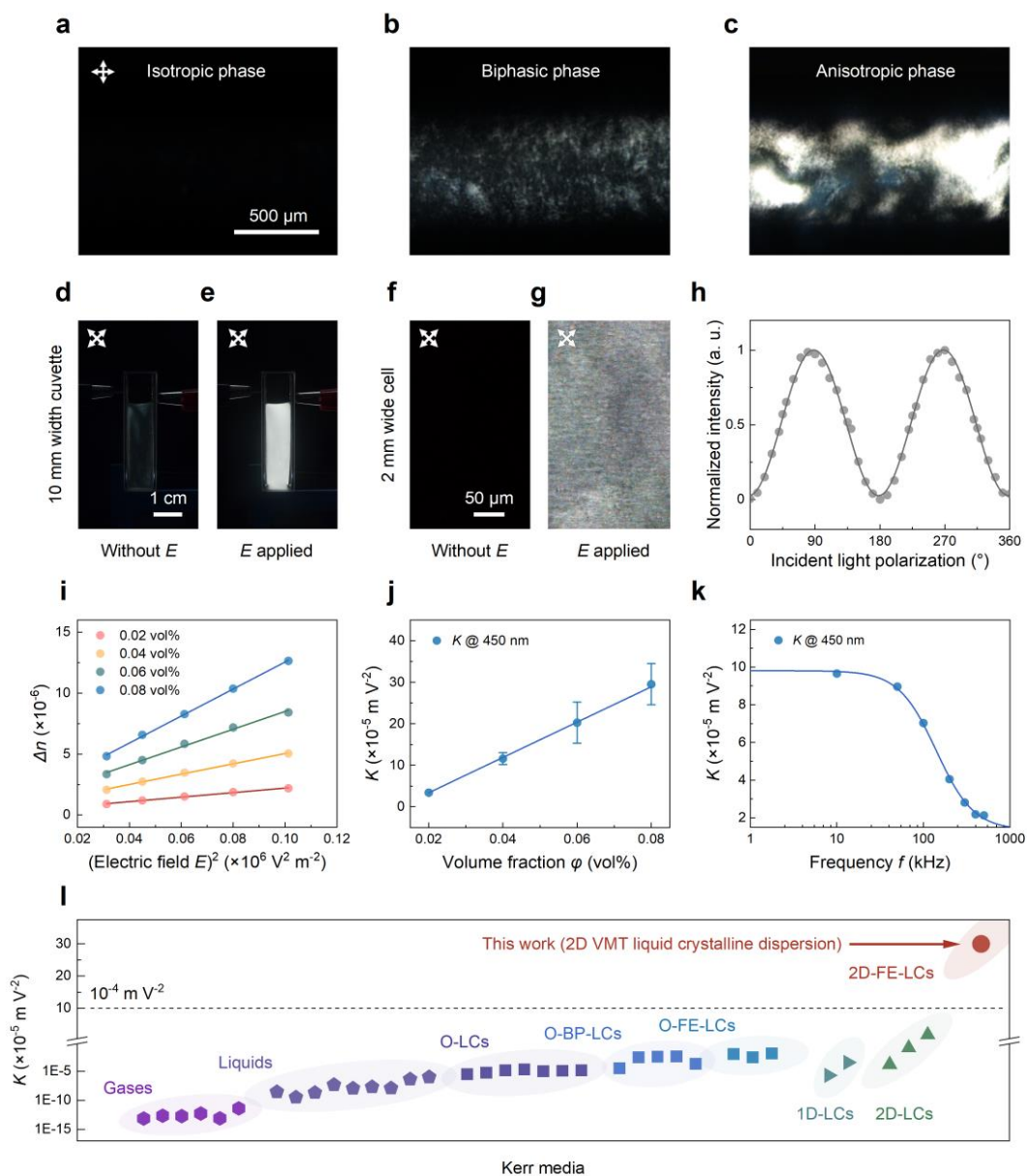

**Fig. 1 | 2D VMT liquid crystalline dispersion and its anomalously large Kerr coefficient. a-c,** Polarized optical microscope images of 2D VMT liquid crystalline dispersion with volume fractions of (**a**) 0.08, (**b**) 0.25, and (**c**) 0.60 vol%. **d-e**, Optical images of a cuvette filled with 2D VMT liquid crystalline dispersion with a volume fraction of 0.08 vol%. The cuvette is placed in front of a white backlight and in-between a pair of crossed polarizers with (**d**) no electric field and (**e**) a transverse electric field of $1.0 \times 10^3$ V m$^{-1}$ ($E$ perpendicular to the direction of optical path). **f-g**, Polarized

optical microscope images of a LC cell containing 2D VMT liquid crystalline dispersion with (**f**) no electric field and (**g**) a transverse electric field of $1.0\times10^4$ V m$^{-1}$. **h**, Normalized intensity of the transmitted light with different incident light polarization. The intensity of the transmitted light with the polarization of incident light perpendicular to the electric field is higher than that parallel to the electric field. **i**, Birefringence $\Delta n$ of 2D VMT liquid crystalline dispersion with volume fractions of 0.02, 0.04, 0.06 and 0.08 vol%. $\Delta n$ varies linearly with the square of the electric field $E$ in fields below $3.2\times10^3$ V m$^{-1}$. **j**, The Kerr coefficient $K$ for 2D VMT liquid crystalline dispersions with volume fractions of 0.02, 0.04, 0.06 and 0.08 vol%. $K$ for 2D VMT liquid crystalline dispersion with a volume fraction of 0.08 vol% is $3.0\times10^{-4}$ m V$^{-2}$. The frequency of all the electric field used in (**d-j**) is 10 kHz. **k**, Frequency dispersion of $K$ in the range of 10 to 1000 kHz. Results in (**h-k**) are collected using a 450 nm laser as an incident light. **l**, Comparison of the measured $K$ values of 2D VMT liquid crystalline dispersion with those of gases, liquids, organic LCs and inorganic nanomaterial LCs reported in the literature, where O-LCs, O-BP-LCs, O-FE-LCs, 1D-LCs, 2D-LCs, and 2D-FE-LCs are classical organic LCs, organic blue phase LCs, organic ferroelectric LCs, inorganic LCs (or LC-like dispersion) based on one-dimensional materials, inorganic LCs (or LC-like dispersion) based on 2D materials, and inorganic LCs (or LC-like dispersion) based on 2D ferroelectric materials, respectively.

In order to fully understand the anomalously large Kerr coefficient, we firstly performed the morphology characterization on 2D VMT liquid crystalline dispersion. Atomic force microscope and transmission electron microscope images revealed that 2D VMT had an average length $\langle D \rangle$ of 2.6 μm (see Fig. 2a-b and Supplementary Fig. 15a), and 70.4% of them had an average thickness of 1.3 nm, which were identified as VMT monolayers (see Fig. 2c and Supplementary Fig. 15b). The proportions of bilayer and trilayer platelets were found to be 20.8% and 6.4%, respectively (Supplementary Fig. 15b). The weighted average height $\langle H \rangle$ was thus determined to be 1.7 nm, giving a large geometrical anisotropy factor γ over 1500, where $\gamma = \frac{\langle D \rangle}{\langle H \rangle}$ for 2D materials. By using the geometrical parameters got above, we showed that the nematic phase transition of 2D VMT liquid crystalline dispersion theoretically occurs at 0.40 vol% (Supplementary Text 5), which is close to that used in our experiment (see Supplementary Fig. 5d). Therefore, the phase transition behavior also supported the results of morphology characterization.

To examine the inherent electric dipole, we attempted ferroelectric characterization using piezoresponse force microscopy (PFM), where 2D VMT was coated on an Au substrate (see Methods and Supplementary Fig. 16). Upon applying an electric field, the amplitude and phase signal of the PFM respectively reflects the local change of morphology caused by the piezoelectric effect and the polarization direction. PFM spectroscopy exhibited a well-defined butterfly-type loop in amplitude, and the phase loop had a hysteresis behavior with a polarization switching of 180° (Fig. 2d). DC poling voltages of -10 V and 10 V were then applied at different positions to write heart and box-in-box patterns on 2D VMT, where the induced polarization is conserved after removal of the DC bias for a ferroelectric material. The PFM images showed clear heart or box-in-box patterns in both amplitude and phase channel after poling, where the phase change of 180° was consistent with the result obtained from PFM spectroscopy (see Fig. 2e-f and Supplementary Fig. 17d, f). We note that both changes in morphology and background signal were negligible (Supplementary Fig. 17a-c, e). A Kelvin probe force microscope image also witnessed the lithographic pattern after poling with a surface potential change >100 mV (Supplementary Fig. 18).

Macroscopic polarization *versus* an external electric field was also measured. A 40 μm-thick 2D VMT laminate film with VMT platelets parallel to each other was prepared by vacuum filtration (Supplementary Fig. 19), and was used to ensure a large polarization signal. A series of macroscopic polarization *versus* voltage (P-V) loops were recorded at room temperature under different driven electric voltages at 0.1 kHz (Fig. 2g). It suggested the existence of electric dipoles in the 2D VMT laminate that can be switched by an external electric field. Note that the observed polarization may include contribution from an unavoidable leakage current, leading to the unsaturated shapes of P-V loops. Such a feature is similar to other relaxor ferroelectric materials[31,32]. Since the contribution from leakage current can be largely suppressed at high frequencies[33,34], a series of P-V loops at high frequencies up to 100 kHz were measured. The switching and hysteresis behavior of P-V loops at high frequency were shown in Supplementary Fig. 20, which gave a polarization of 0.03 μC cm$^{-2}$ and confirmed the essential contribution from electric dipoles. Thus, we presumably ascribed the polarization to an inherent electric dipole. To reveal the origin of the inherent electric dipole, we performed first principle calculations on a monolayer VMT (see Methods). Supplementary Text 6 and Supplementary Fig. 21 showed that the vacancies induced by the elemental substitution (e.g., by Al and Fe) in monolayer VMT is a possible origin for the inherent electric dipole and resultant ferroelectricity. Such a mechanism relies on the migration of proton from one side of the $MgO_2$ sheet to the other side. In this case, 2D VMT possessed a flipped polarization in both in-plane and out-of-plane direction, and the in-plane polarization is larger than the out-of-plane polarization, which agreed with our experimental results (Fig. 1h, Supplementary Fig. 11, and Supplementary Fig. 22).

Besides, it is worth noting that the lyotropic liquid crystalline dispersion with ferroelectric building units (i.e., 2D VMT platelets in this work) differ from the generally understood thermotropic ferroelectric LCs. One can observe the polar ordered domains and their transition controlled by a DC electric field in thermotropic ferroelectric LCs[15,35,36], while there was no similar phenomenon for 2D VMT liquid crystalline dispersion until the DC electrophoresis took place (Supplementary Fig. 23). In theory, interactions between molecules in ferroelectric LCs typically involve three fundamental mechanisms, i.e., charge-charge interaction $V_{cc}$, dipole-charge interaction $V_{cd}$, and dipole-dipole interaction $V_{dd}$. The strength of these interactions varies with the distance between adjacent molecules, respectively following inverse relationships of the first, second, and third orders of the distance of $r$, namely, $V_{cc} \propto r^{-1}$, $V_{cd} \propto r^{-2}$, and $V_{dd} \propto r^{-3}$. When molecules possess strong surface charges (Supplementary Fig. 4c), charge-charge repulsion becomes the dominant force, pushing molecules away from each other to several hundreds of nanometers (726 nm for 2D VMT liquid crystalline dispersion, based on Derjaguin–Landau–Verwey–Overbeek theory, see Supplementary Text 7, and Supplementary Fig. 24), thereby weakening dipole-charge interaction and dipole-dipole interaction to a negligible level. Thus, the individual ferroelectric nature of LC building units directly contributes to the increase of electro-optical sensitivity (i.e., the Kerr coefficient), rather than forming collective polar ordered domains in the case of thermotropic ferroelectric LCs.

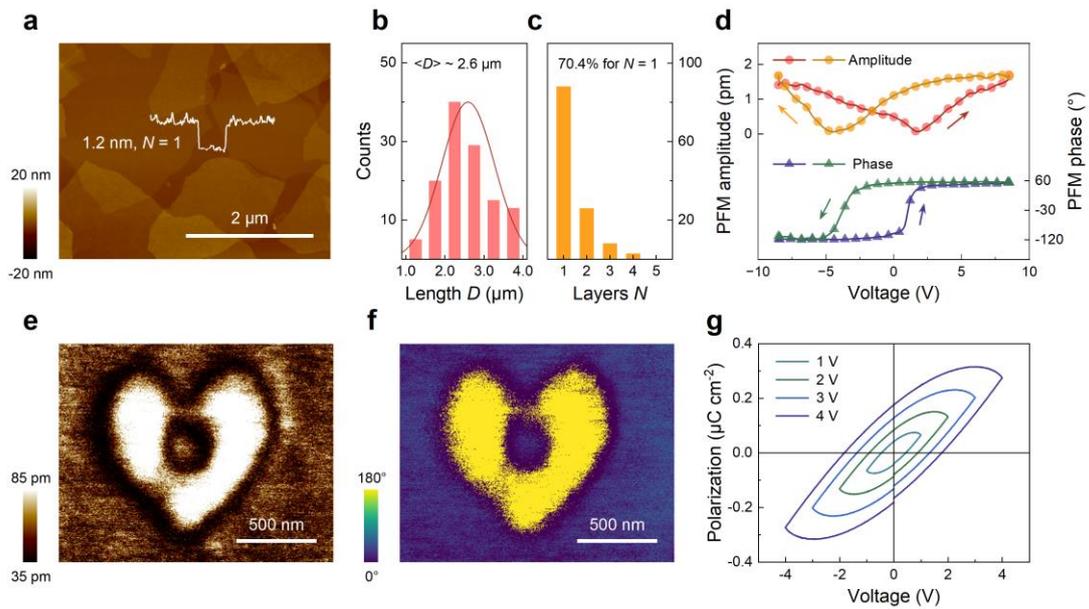

**Fig. 2 | 2D VMT and evidence of ferroelectricity. a**, Atomic force microscopy image of 2D VMT. The height profile shows a height difference of 1.2 nm between the substrate and the 2D VMT platelets. The platelets are assigned as VMT monolayers, i.e. $N = 1$ where $N$ is the number of layers. **b-c**, Statistics of (**b**) the length $D$ and (**c**) the number of layers of the 2D VMT. $\langle D \rangle$ is determined to be 2.6 μm and 70.4% of 2D VMT is monolayer. Bilayer ($N = 2$) and trilayer ($N = 3$) platelets account for 20.8% and 6.4%, respectively. **d**, PFM amplitude and phase *versus* voltage hysteresis loops of 2D VMT. **e-f**, (**e**) PFM amplitude and (**f**) PFM phase maps of 2D VMT scanned after

writing a heart pattern by applying a DC bias of -10 V and 10 V. **g**, P-V hysteresis loops of a 2D VMT laminate film under different voltage biases at room temperature.

Nevertheless, these results are of importance to examine the giant Kerr coefficient of 2D VMT liquid crystalline dispersion. Supplementary Text 8 and 9 gave their insights on theoretical determination of optical anisotropy factor $\Delta g$ and excess electrical polarizability anisotropy $\Delta \alpha$ by using the geometrical parameters we got in Fig. 2a-b and Supplementary Fig. 15b (average length $\langle D \rangle$ of 2.6 μm, weight average height $\langle H \rangle$ of 1.7 nm, and a geometrical anisotropy factor $\gamma$ of 1500). The $\Delta g$ and $\Delta \alpha$ were determined to be $-2.2 \times 10^{-12}$ $C^2$ $J^{-1}$ $m^{-1}$ and $-8.1 \times 10^{-27}$ F $m^2$, respectively, which agreed well with the experimental results ($-2.3 \times 10^{-12}$ $C^2$ $J^{-1}$ $m^{-1}$ and $-9.3 \times 10^{-27}$ F $m^2$). By combining these geometrical parameters with the polarization of 0.03 μC $cm^{-2}$, we calculated the inherent electric dipole $|\mu|$ of 2D VMT liquid crystalline dispersion independently. Supplementary Text 10 presented a $|\mu|$ of $1.8 \times 10^{-23}$ C m, being close to $1.7 \times 10^{-23}$ C m as well. Note that, Supplementary Text 11 showed that all the three aforementioned critical parameters ($|\Delta g|$, $|\mu|$ and $|\Delta \alpha|$) that determine the Kerr coefficient were monotonically increased with an increasing geometrical anisotropy factor of the LC units. For 2D VMT possessing an intrinsic ferroelectric response with a large geometrical anisotropy factor $\gamma$ of >1500 and an intrinsic polarization $P$ of 0.03 μC $cm^{-2}$, the contribution of $|\mu|$, i.e., $\left(\frac{\mu}{k_B T}\right)^2$ of about $1.9 \times 10^{-5}$ $m^2$ $V^{-2}$ was much greater than that of $|\Delta \alpha|$, i.e., $\frac{\Delta \alpha}{k_B T}$ of about $2.3 \times 10^{-6}$ $m^2$ $V^{-2}$ according to our calculation. In such case, an approximate optimization gave a generalized formula of $K \propto P^2 \gamma^4$ that showed a strong positive correlation of Kerr coefficient with an intrinsic polarization $P$ and a geometrical anisotropy factor $\gamma$ (Supplementary Text 11). To clearly explain this issue, we further verified the derived relationship by experiments. First, 2D VMT liquid crystalline dispersion with a volume fraction of 0.08 vol% was bath sonicated for 30, 60, and 120 minutes, where Supplementary Fig. 25 showed that the length of 2D VMT decreased with increasing durations of sonication, and the average lengths of them were determined to be 1.4, 1.0, and 0.7 μm. Since the as-exfoliated 2D VMT are monolayer dominant, a weight average height of 1.7 nm was taken into calculation, which gave their average geometrical anisotropy factors of 841, 561, and 401. Supplementary Fig. 26 showed the results of Kerr experiments. The four liquid crystalline dispersions in which 2D VMT had average geometrical anisotropy factors of 1525 (without sonication), 841, 561, and 401 showed Kerr coefficients of $3.0\times10^{-4}$, $3.4\times10^{-5}$ m V$^{-2}$, $4.8\times10^{-6}$ m V$^{-2}$, and $1.1\times10^{-6}$ m V$^{-2}$, respectively. Noteworthy, the fitting curve gave a relationship of $K \propto \gamma^{4.2}$, being close to our derivation that $K \propto \gamma^4$. Besides, the Kerr coefficients of 2D VMT liquid crystalline dispersion showed a negative quadratic dependence with the temperature as $K \propto T^{-2.2}$ (Supplementary Fig. 27). In principle, the relationship of $K \propto T^{-2}$ only occurred in the case that the contribution of $|\mu|$ dominates. Hence, we proposed that an extremely large geometrical anisotropy and an intrinsic polarization of 2D VMT jointly increased the sensitivity of the electro-optical Kerr effect to a new record of $3.0\times10^{-4}$ m V$^{-2}$, which is one order of magnitude higher than all known media.

So far, producing a LC based large-scale display for outdoor use remains challenging, mainly due to its high operational electric field of >10$^6$ V m$^{-1}$, as well as the concern about electrophoresis of ion species in R&D LC systems, high energy consumption, complicated/costly preparation of transparent electrodes with a micrometer-scale space[3,9,10]. A giant Kerr coefficient makes it possible to fabricate such a display with low energy consumption and high operational durability, where the operational electric field can be theoretically dropped from 10$^6$ V m$^{-1}$ for organic LCs to 10$^2$~10$^4$ V m$^{-1}$ for the 2D VMT liquid crystalline dispersion. Such a low operational field consequently helps overcome the aforementioned problems. We designed and fabricated 2D VMT liquid crystalline dispersion based pixels with a size of 1.4 inch, which had two working modes with its backlight supplied from a light emitting diode screen for night use, or reflection of a solar beam for daytime use (Fig. 3a). The pixel was comprised of two electrodes separated in centimeter-scale space, supplying an electric field in the range of 10$^2$~10$^4$ V m$^{-1}$. Since the direction of the electric field was perpendicular to the light path in the 2D VMT liquid crystalline dispersion pixel, transparent electrodes, such as indium tin oxide or fluorine doped tin oxide, were not required, permitting the use of cost-effective non-transparent metals, such as copper, as alternative electrodes. In the meantime, because of its wide optical band gap and corresponding absorption edge of > 400 nm, 2D VMT liquid crystalline dispersion with a volume fraction of 0.04 vol% had an average transmittance of 88% for the visible light (Supplementary Fig. 28). As a result, we loaded such dispersion into the pixel that can be scaled up to inch level. With the use of color filters (Fig. 3b) or monochromic lights with different wavelengths (see Fig. 3c and Supplementary Video 1), the pixel displayed red, green and blue colors.

The 2D VMT liquid crystalline dispersion pixels can be assembled into an array with display functions. For instance, we designed and fabricated a prototypical displayable billboard for potential interactive large-screen use in smart architectures (see Methods and Supplementary Fig. 29). A hash addressing algorithm was developed to determine the specific pixels that will be turned on in the billboard by controlling corresponding relays (see Supplementary Fig. 30 and Supplementary Table 3). The displayable billboard can be remotely controlled by a smartphone, and showed letters "T", "H" and "U" (see Fig. 3d and Supplementary Video 2). By integrating it with a hand-tracking module, it can automatically display numbers such as "1", "2", and "3" after capturing the gestures of human hand (see Fig. 3e and Supplementary Video 3). The array also had some other attractive performance parameters, such as high uniformity, low power consumption and negligible ultraviolet degradation. For instance, a display with 97% consistency among all pixels was found by luminance tests (Supplementary Fig. 31). For an array that the backlight was supplied from the reflection of a solar beam, the power consumption was approximately 123 W m$^{-2}$ and was reduced by 35% compared with that using light emitting diode screen as a backlight (~189 W m$^{-2}$) (see Supplementary Text 12 and Supplementary Table 4). The energy consumption of the 2D VMT liquid crystalline dispersion display (100-200 W m$^{-2}$) was comparable to commercially available LC display techniques[37]. We performed a long-term operational test of the stability of 2D VMT liquid crystalline dispersion display, which gave a small decay of <1% for both outdoor (after 1000 hours under sunlight) and indoor tests (after 8000 hours) (Fig. 3f). In contrast, organic LCs are likely to undergo photo-degradation after exposure to UV irradiations for hours[38,39], making its outdoor use challenging. In addition, a transient test was performed to examine the dynamic process and the response time of the device. Supplementary Fig. 32 and Supplementary Text 13 collectively showed that both the rising and falling edge agreed well with theoretical exponential processes, when polydispersity is taken into account. The characteristic time of rising edge and that of falling edge were determined to be 0.35 s and 2.05 s, respectively. Compared with devices based on commercial LCs with a response time of a few milliseconds or even smaller, the response time indeed exists a room for improvement, but it is still satisfactory for some applications, such as advertisement presentation and smart windows. Finally, we compared the 2D VMT liquid crystalline dispersion display with other electrochromic techniques in a radar plot (see Methods, Fig. 3g and Supplementary Table 5). It can be seen that 2D VMT liquid crystalline dispersion display had advantages in small-field operation, outdoor stability, optical density and contrast ratio. Above all, the power-saving, ultraviolet-stable, and integration compatible 2D VMT liquid crystalline dispersion devices offer clear opportunities for outdoor LC optics in the future.

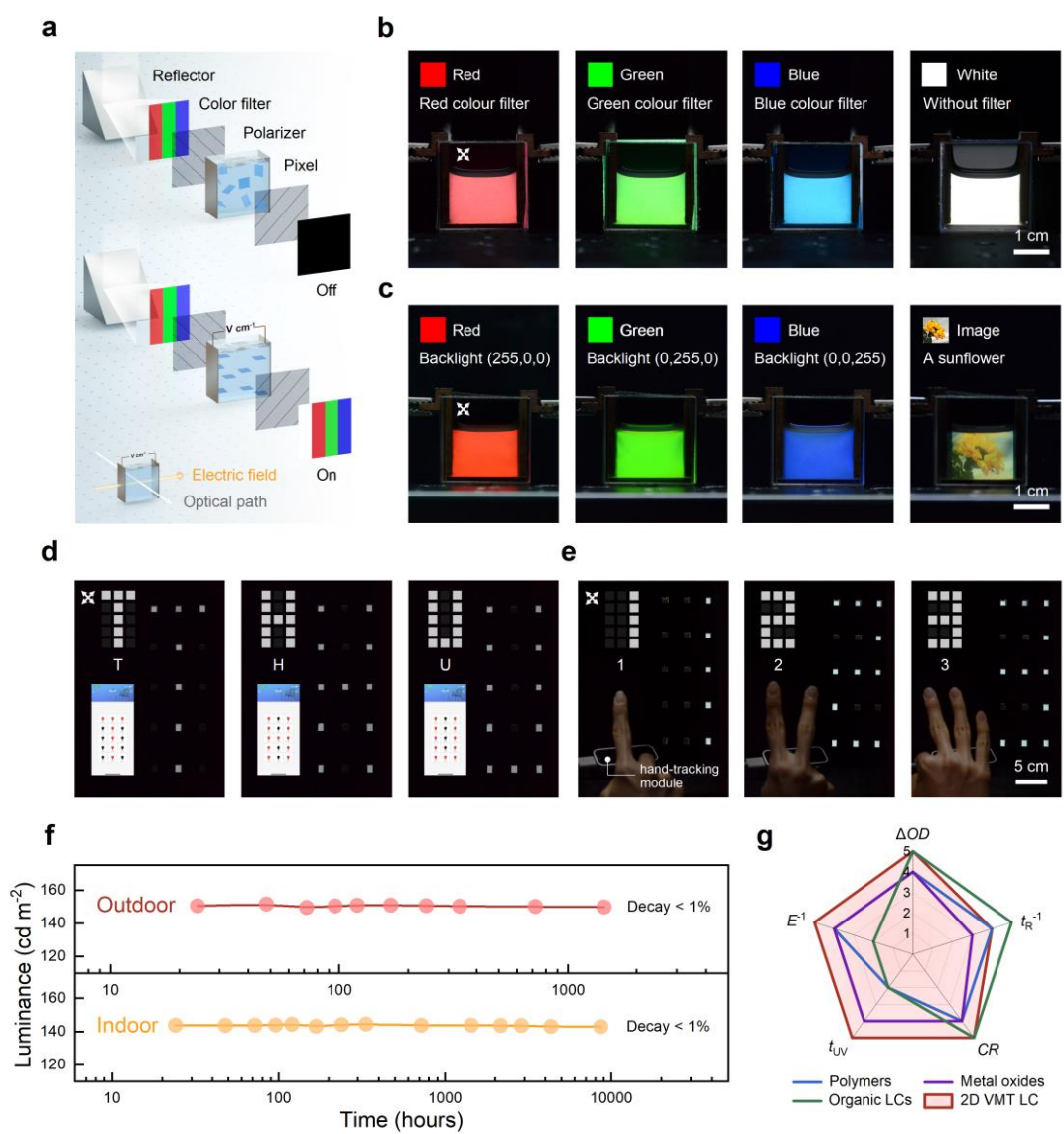

**Fig. 3 | Proof-of-concept devices using 2D VMT liquid crystalline dispersion. a**, Schematic of a backlight-free displayable pixel using a reflector. The pixel appears black without an electric field. An electric field of $10^2$~$10^4$ V m$^{-1}$ is used to turn on the device. Color filters are used to get red, green and blue colors. The bottom left inset presents the device configuration with the electric field direction perpendicular to the optical path. **b**, Optical images of a backlight-free displayable pixel showing red, green, blue, and white colors with an applied electric field of $4.0 \times 10^3$ V m$^{-1}$. **c**, Optical images of a displayable pixel with a backlight module showing red, green, blue colors and a sunflower picture with an applied electric field of $4.0 \times 10^3$ V m$^{-1}$. **d**, Pixel array remotely controlled by a software. The array shows the letters "T", "H" and "U" with an applied

electric field of 8.0×10³ V m⁻¹. **e**, Pixel array remotely controlled by gestures of the controller. A hand-tracking module is equipped to capture the gesture. The array shows the numbers "1", "2" and "3" with an applied electric field of $8.0×10^3$ V m⁻¹. The frequency of all the electric field used is set as 10 kHz in (**b-e**). **f**, Luminance of a 2D VMT liquid crystalline dispersion pixel during long-term outdoor and indoor stability tests. The decay is less than 1% in both cases. **g**, Comprehensive comparison of 2D VMT liquid crystalline dispersion device with those reported electrochromic techniques reported in the literature. Five parameters are used: outdoor stability (represented by life in ultraviolet irradiation, $t_{UV}$), manipulating field ($E^{-1}$), optical density ($\Delta OD$), response time ($t_R^{-1}$), and contrast ratio ($CR$).

We have discovered a 2D VMT liquid crystalline dispersion with an anomalously large electro-optical Kerr coefficient of $3.0×10^{-4}$ m V⁻². Our work provides inspiring evidence on 2D ferroelectrics in layered natural minerals, which sheds lights on the scalable production of van der Waals ferroelectric materials. These results also indicate opportunities to design advanced inorganic LCs or LC-like systems with an ultra-high electro-optical performance by preparing 2D materials from bulk ferroelectric materials.

## Methods

**Preparation of 2D VMT liquid crystalline dispersion and characterization**

Vermiculite (VMT, sizes: 2-3 mm, Sigma-Aldrich, USA), sodium chloride (NaCl, purity >99.5%, Shanghai Macklin Biochemical Co., Ltd., China) and lithium chloride (LiCl, purity >99.0%, Shanghai Macklin Biochemical Co., Ltd., China) were used without any treatment. The 2D VMT liquid crystalline dispersion was obtained following a two-step ion-exchange method. The VMT was immersed in 100 mL of saturated NaCl solution (26.5 wt%) and stirred on a hot stage at 80 °C for 24 h. The solution was collected and washed with deionized water repeatedly to remove residual salts. The $Na^+$ exchanged VMT precursor was immersed in 100 mL 2 M LiCl solution and stirred on a hot stage at 80 °C for another 24 h. The products were repeatedly washed with deionized water until the ionic strength went below $10^{-4}$. The unexfoliated VMT was removed by ultracentrifugation. The morphology of the 2D VMT was investigated by atomic force microscopy (AFM, tapping mode, Cyper ES, Oxford Instruments, USA) and transmission electron microscopy (TEM, 120 kV, Spirit T12, FEI, USA). The stability of the 2D VMT liquid crystalline dispersion was examined by a zeta potential analyzer (Zetasizer Nano-ZS90, Malvern, UK). Small angle X-ray scattering data was obtained using a Nanostar SAXS system, Bruker, Germany. The rheological behavior was studied using a rheometer (MCR-302, Anton Paar, Austria).

**PFM and polarization measurements**

2D VMT was coated on a conducting Au substrate for PFM measurement using a Langmuir−Blodgett method. The 2D VMT dispersion was vacuum-filtrated to remove the deionized water, and 2D VMT remaining on the filtration substrate was re-dispersed in a mixed solvent of methanol and chloroform with a ratio of 1:1. Then, it was carefully dropped onto the surface of a water bath, in which an Au substrate had been submerged. Two baffles on opposite sides of the bath were steadily moved towards each other to achieve a tight arrangement of 2D VMT on the water surface until the surface pressure reached 10 mN $m^{-1}$. The Au substrate was pulled up at a rate of 1 mm $min^{-1}$, where 2D VMT adhered to its surface. PFM measurements were carried out by atomic force microscopy (Cyper ES, Oxford Instruments, USA) in a Dual AC Resonance Tracking mode. PFM amplitude and phase loops were recorded using the spectroscopy channel. The heart and box-in-box patterns were written by applying DC poling voltages of -10 V and 10 V at different positions with a sweep rate of 1.92 Hz. A conducting tip with a resonance frequency of 350-400 kHz was driven by a DC voltage in the range of 0.8-1.5 V to scan the sample after the writing process. An additional voltage of 2.2-2.8 V was applied to the tip to eliminate the surface electrostatic effect of 2D VMT. 50 mL of a 2D VMT Kerr dispersion with a volume fraction of 0.04 vol% was used to prepare an assembled 2D VMT laminate film by vacuum filtration to characterize the macroscopic polarization. Polarization *versus* voltage hysteresis loops were obtained using a ferroelectric tester (Precision Premier II, Radiant Technologies, USA). The testing frequencies were 0.1 kHz and 100 kHz.

**Ferroelectricity calculations**

The density-function-theory calculations were performed with the use of projected augmented wave potential implemented Vienna Ab initio Simulation Package (VASP 5.4.4) code[40-42]. The exchange-correlation interactions were described within the generalized gradient approximation using the Perdew-Burke-Ernzerhof form[43]. The DFT-D3 method with Becke-Johnson damping function was used to describe the van der Waals interactions[44,45]. Cut-off energy of 520 eV was set for the plane-wave basis, and the Brillouin zone was sampled with 6×4×1 k-points using the Monkhorst-Pack scheme[46]. The shape and volume of unit cell were fully optimized with the convergence criteria being $1\times10^{-6}$ eV for the energy difference in the electronic self-consistent iteration and $1\times10^{-3}$ eV Å$^{-1}$ for the residual force on all atoms. The vacuum layers were set as 23 Å so that the interaction between two neighboring bilayer can be neglected. The Berry phase method was used to compute electrical polarizations[47], and the climbing image nudged elastic band method was adopted to compute the ferroelectric switching pathway[48].

**Electro-optical measurements**

A quartz cuvette, with a square cross-section and 10 mm between opposite internal faces (10 mm × 10 mm × 45 mm), was filled with the 2D VMT liquid crystalline dispersion and used for electro-optical measurements. Two parallel copper plates were placed on the other internal faces of the cuvette and were used as counter electrodes to supply the electric field perpendicular to the optical path. A generator (AFG 3102C,

Tektronix Inc., USA) and an amplifier (ATA-2082, Aigtek Co. Ltd., China) were combined to provide an electric field with frequencies of 10-1000 kHz. A polarizer (GL10-A, Thorlabs Inc., USA) was placed in the incident light path. Optical images were taken by a polarized optical microscope (Imager A2m, Carl Zeiss, Germany) or a Nikon D7000 digital camera with a lens of AF-S DX Nikkor 18-140 mm f/3.5–5.6G ED VR. A 450 nm laser was used as the incident light. A power meter (PM 200, Thorlabs Inc., USA) or a polarimeter (PAX1000, Thorlabs Inc., USA) was put in the output optical path as a detector. For polarization-dependent transmittance, the polarizer was rotated for 360 degrees, and the intensity of transmitted light was recorded by a power meter. For quantitative measurement of the electro-birefringence $\Delta n$, the polarization direction was set at 45 degrees to the direction of the electric field. The polarization signal recorded by the polarimeter can be converted to $\Delta n$ using $\Delta n = \frac{\lambda}{2\pi L}\left\{arctan[tan(\theta) \times tan(\eta)] + arctan\left[\frac{tan(\eta)}{tan(\theta)}\right]\right\}$, where $\lambda$ is the wavelength, $L$ the optical path, $\theta$ the azimuth angle, and $\eta$ the ellipticity. A detailed derivation was given in Supplementary Text 1.

**Fabrication of prototypical devices**

A square quartz cuvette with a display area of 1.4 inches was fabricated to be a proof-of-concept displayable pixel. The polarizers and electrodes were set as mentioned earlier in "Electro-optical measurement" section. For a backlight-free pixel, an aluminum mirror was put behind it. We placed red, green or blue filters between the mirror and the pixel to generate colors from the natural light. A screen displaying standard red with a RGB color code of (255, 0, 0), green (0, 255, 0), blue (0, 0, 255) and colored sunflower pictures was used as a backlight when needed. An electric field of $4\times10^3$ V m$^{-1}$ was applied to turn on the pixel, which is black without an external electric field. To integrate the pixels, we designed a printed circuit board, where 15 pixels were assembled on it, as shown in Supplementary Fig. 29. An open-source Arduino chip (Arduino Mega, Creative Commons) was used to dynamically control the relay array, to which all the pixels were connected. A home-developed hash addressing algorithm was pre-written on the Arduino chip, which was responsible for finding the relays to be turned on when receiving a signal. A wireless communication module remotely controlled by software was connected to the billboard if needed. A hand-tracking module (Leap Motion Controller, Ultraleap Limited, USA) was used to capture the gestures of a human hand. For both cases, the electric field was pre-set as 0, $4\times10^3$ or $8\times10^3$ V m$^{-1}$. The electric field was applied when the controller clicked the pixel icons shown in the smartphone software (Blinker, Diandeng technology, China) or the hand-tracking module capture the gesture of the controller. All images were taken by a Nikon D7000 digital camera with a lens of AF-S DX Nikkor 18–140 mm f/3.5–5.6G ED VR.

**Evaluation of device performance**

Operational stability was monitored by recording the luminance of the pixel for long-term operation. The luminance was measured by a spectroradiometer (SpectraDuo PR-680, Photo Research Inc., USA). For an indoor test, we kept the 2D VMT liquid crystalline dispersion in a dark place. For an outdoor test, we exposed it to sunlight, where we estimated an exposure time of 8 hours per day. The response time ($t_R$) was examined by a transient test and the rising edge and falling edge both follow an exponential process. Here, we used the characteristic time of rising edge to represent the response time $t_R$. The optical density ($\Delta OD$) was calculated according to $\Delta OD = log \frac{T_{on}(\lambda_{max})}{T_{off}(\lambda_{max})}$, where $T_{on}(\lambda_{max})$ and $T_{off}(\lambda_{max})$ are respectively transmittance of the pixel at on and off states at the peak wavelength. The Michelson contrast was used following $CR = \frac{T_{on}^{int} - T_{off}^{int}}{T_{on}^{int} + T_{off}^{int}}$. Here, $T_{on}^{int}$ and $T_{off}^{int}$ are integrated transmittance in visible spectral region at off and on states.

**Data availability**

The data that support the findings of this study are available within the paper and the Supplementary Information. Other relevant data are available from the corresponding authors on reasonable request. Source data are provided with this paper.

## References for Methods

## Acknowledgements

This work is supported by the National Natural Science Foundation of China (Nos. 51920105002, 52125309, 52273311, 51991343, 51991340, and 52188101), Guangdong Innovative and Entrepreneurial Research Team Program (No.

2017ZT07C341), and Shenzhen Basic Research Project (Nos. WDZC20200819095319002, JCYJ20190809180605522, and JCYJ20220818100806014). The authors also acknowledge Yue Gu, Rui Gong, Siyuan Tian, Lixin Dai, Tianshu Lan, Yujie Sun, Zhongyue Wang, Shengkai Chang, Hefei Xi and Xiaolong Zou for their inputs.

## Author contributions

Z. H., Z. Z., B. D. and B. L. conceived the idea. Z. H., Y. X., and J. L. prepared 2D VMT and performed characterization. R. Z., K. W. and T. M. contributed to the PFM tests. B. D., G. Z. and C. R. investigated the macroscopic polarization. L. Y. and M. W. conducted the ferroelectricity calculations. Z. H., Y. X., J. L., Y. H., Z. Z. and B. D. investigated the electro-optical birefringence effect and performed theoretical analyses. Z. Z., B. D. and Z. H. fabricated the devices and evaluated the performance. All authors discussed and analyzed these results in detail. Z. H., Z. Z., B. D. and B. L. wrote the manuscript assisted by other authors. All authors reviewed and revised the manuscript.

## Competing interests

B. L., Z. Z., B. D., and Z. H. are inventors of a patent related to this work. The patent is held by Tsinghua Shenzhen International Graduate School, Tsinghua University.

## Additional information

Supplementary Information with Supplementary Texts 1-13, Supplementary Figures 1-32, Supplementary Tables 1-5, Captions for Supplementary Videos 1-3, and Supplementary References 1-50 is available. Other Supplementary Information for this manuscript includes Supplementary Videos 1-3.